\documentclass[a4paper,pre,twocolumn,amsmath,amssymb,longbibliography]{revtex4-1}
\usepackage{bm}
\usepackage{graphicx,color,soul}
\usepackage[colorlinks,linkcolor=magenta,citecolor=magenta]{hyperref}

\newcommand{\dd}{\mathrm{d}}

\newcommand{\pd}[2]{\frac{\partial #1}{\partial #2}}

\newcommand{\Int}[1]{\int\dd #1\;}
\newcommand{\IInt}[3]{\int_{#2}^{#3}\dd #1\;}
\renewcommand{\vec}[1]{\mathbf #1}

\newcommand{\al}{\alpha}
\newcommand{\gam}{\gamma}

\newcommand{\kap}{\kappa}
\newcommand{\lam}{\lambda}

\newcommand{\sig}{\sigma}
\newcommand{\msig}{\bm\sigma}

\newcommand{\dgam}{\dot\gam}

\AtBeginDocument{%
    \newwrite\bibnotes
    \def\bibnotesext{Notes.bib}
    \immediate\openout\bibnotes=\jobname\bibnotesext
    \immediate\write\bibnotes{@CONTROL{REVTEX41Control}}
    \immediate\write\bibnotes{@CONTROL{%
    apsrev41Control,author="08",editor="1",pages="1",title="0",year="1"}}
     \if@filesw
     \immediate\write\@auxout{\string\citation{apsrev41Control}}%
    \fi
}%

\begin{document}

\title{Steady inhomogeneous shear flows as mechanical phase transitions}

\author{Thomas Speck}
\affiliation{Institute for Theoretical Physics IV, University of Stuttgart, Heisenbergstr. 3, 70569 Stuttgart, Germany}


\begin{abstract}
  Inhomogeneous flows and shear banding are of interest for a range of applications but have been eluding a comprehensive theoretical understanding, mostly due to the lack of a framework comparable to equilibrium statistical mechanics. Here we revisit models of fluids that reach a stationary state obeying mechanical equilibrium. Starting from a non-local constitutive relation, we apply the idea of a ``mechanical phase transition'' and map the constitutive relation onto a dynamical system through an integrating factor. We illustrate this framework for two applications: shear banding in strongly thinning complex fluids and the coexistence of a solid with its sheared melt. Our results contribute to the growing body of work following a mechanical route to describe inhomogeneous systems away from thermal equilibrium.
\end{abstract}

\maketitle


\section{Introduction}

One of the most basic mechanical properties of matter is how it responds to deformation. While solids deform elastically and withstand (small) strains, fluids start to flow immediately. This behavior can be captured by measuring the externally applied shear stress $\sig$ together with the rate of deformation, typically the strain rate $\dgam$. For simple Newtonian fluids, these are proportional to each other with the proportionality given by the viscosity, which is a material property. On the other hand, complex fluids such as colloidal suspensions and polymer solutions exhibit pronounced non-linear behavior~\cite{larson99,larson15,royall24} including shear thinning and even shear thickening~\cite{morris20}. Much work has been done to elucidate the microscopic origins of these phenomena and to link rheology with structural changes in the statistical arrangement of the solvated constituents.

A rather intriguing manifestation of non-linear behavior is the emergence of inhomogeneous flows, in which regions with different strain rates coexist. These \emph{shear bands} have been observed in a range of complex fluids~\cite{moller08,chikkadi14} and numerical models~\cite{greco97,varnik03,chaudhuri12,martens12,nicot23}, and have been the subject of a number of reviews~\cite{olmsted08,schall10,fielding14,divoux16}. In (athermal) disordered solids, shear bands form through the localization of plastic deformations~\cite{wisitsorasak17} and thus indicate a material failure that has prompted extensive investigations. Theories have linked shear bands in these materials to critical behavior~\cite{parisi17,ozawa18}.

In thermal systems, different phases can coexist at equilibrium. Statistical physics provides two conditions, the equality of pressure and chemical potential, that together allow to unambiguously determine the bulk phases (sufficiently far from the interfacial region). Given a bulk equation of state (including a region in which homogeneous states are unstable), the procedure is known as equal-area Maxwell construction. Although inhomogeneous steady flows share the property of coexisting domains, they are driven far from thermal equilibrium and thus these concepts from equilibrium statistical physics cannot be applied directly. A generalization of the Maxwell construction from a purely mechanical perspective has been first discussed by Aifantis and Serrin~\cite{aifantis83,zbib92}. More recently, determining phase coexistence far from equilibrium has been tackled for active (i.e., self-propelled) Brownian particles following essentially equivalent arguments~\cite{solon18,solon18a,tjhung18,speck21a,omar23,langford24}. This framework has been extended to active mixtures~\cite{chiu24a} and non-reciprocal interactions~\cite{saha24,greve24}. In a nutshell, an integrating factor together with the condition of a uniform chemical potential (or stress) is used to remove ``non-equilibrium'' (i.e., not following from a free energy functional) interfacial terms, shifting their impact to an effective bulk term. Here we revisit the theory of inhomogeneous flows and show that the same framework can be applied, unambiguously predicting the coexisting strain rates from a non-local constitutive relation together with the condition of mechanical equilibrium even though the equal-area construction does not apply.

\begin{figure}[b!]
  \centering
  \includegraphics{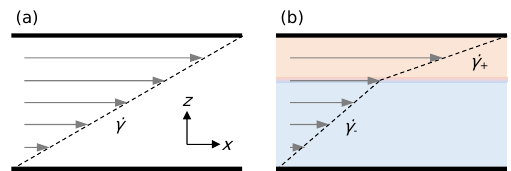}
  \caption{Sheared complex fluids. (a)~Uniform flow profile (gray arrows) with a constant strain rate $\dgam$. (b)~Inhomogeneous flow profile with two regions in which the strain rate is approximately constant separated by an interfacial region.}
  \label{fig:sketch}
\end{figure}


\section{Theory}

\subsection{Constitutive relation}

We are interested in a system bounded by two plates that is sheared in $x$-direction and possibly becomes inhomogeneous perpendicular to the plates in $z$-direction (Fig.~\ref{fig:sketch}). We assume a steady state so that the (macroscopic) stress tensor $\msig=\msig(z)$ obeys $\nabla\cdot\msig=\partial_z\sig=0$ with shear stress $\sig=\sig_{xz}$. This implies that the shear stress is constant $\sig(z)=\bar\sig$ throughout the system.

Now suppose the system is a (complex) fluid confined between the plates that has a laminar flow profile $\vec u=u\vec e_x$ due to the shearing. If the flow profile is uniform, $u(z)=\dgam z$ with constant strain rate $\dgam$, the bulk stress $\sig_0=\sig_0(\dgam)$ is a function of the strain rate $\dgam$ alone. If the flow profile $\dgam=\dgam(z)$ becomes spatially non-uniform, \emph{i.e.} different regions of the fluid flow with different strain rates, then such a simple function $\sig_0(\dgam)$ cannot fulfill the mechanical equilibrium condition anymore. However, in a non-uniform system quite generally the stress should depend also on spatial derivatives of the strain rate. Demanding that the stress changes sign as $\dgam\to-\dgam$ and is invariant with respect to inversion $z\to-z$, to lowest order one obtains the non-local constitutive relation
\begin{equation}
  \sig = \sig_0 - \kap\partial_z^2\dgam
  \label{eq:cr}
\end{equation}
with coefficient $\kap=\kap(\dgam)$ that can depend on the local strain rate~\cite{spenley96,dhont99,olmsted99}. Assuming that the microstructure, to which the stress can be related, saturates at large strain rates, the stress no longer depends on the spatial derivative and $\kap(\dgam\to\infty)\to 0$ should vanish in this limit.

A uniform state becomes mechanically unstable if $\partial_{\dgam}\sig_0<0$ since the system can reduce stress through increasing the strain rate (i.e., flowing faster)~\cite{yerushalmi70}. At fixed applied strain rate, the system escapes the instability through splitting into two (or more) regions with different strain rates outside the unstable range of strain rates. The stress $\bar\sig$ is then determined by the interface separating these two regions. Let us assume two extended regions with constant strain rates $\dgam_\pm$ separated by an interfacial region in which $\dgam(z)$ varies quickly. We pick two points at $z_-$ and $z_+$ so that $\dgam_\pm\equiv\dgam(z_\pm)$ [cf. Fig.~\ref{fig:sketch}(b)]. Rearranging the constitutive equation~\eqref{eq:cr} followed by integration yields
\begin{equation}
  \IInt{\dgam}{\dgam_-}{\dgam_+} \frac{\bar\sig-\sig_0(\dgam)}{\kap(\dgam)} = 0
  \label{eq:maxwell}
\end{equation}
since gradients $\partial_z\dgam\to0$ vanish in the bulk regions. This is the modified Maxwell construction of Refs.~\cite{dhont99,olmsted99} determining the shear stress $\bar\sig$. We will now explore an alternative route to $\bar\sig$ closer to the ideas of Aifantis and Serrin.

\subsection{Integrating factor}

While we could simply ``translate'' the result of, e.g., Ref.~\cite{speck21a} for the coexistence of active Brownian particles, it is instructive to develop the theory from the perspective of a dynamic system. A constant shear stress $\sig(z)=\bar\sig$ is a necessary condition for a steady state, but it does not yet determine the strain profile in an inhomogeneous system. In a thermal system at equilibrium, the equality of chemical potential provides a second (and sufficient) condition that ensures the free energy is minimal. For a constitutive relation of the form in Eq.~\eqref{eq:cr}, we now construct a function $E(z)$ that plays a similar role. To this end, we introduce an integrating factor $\partial_zv$ with some, at this point unknown, function $v=v(\dgam)$ so that $\sig\partial_zv=-\partial_z\psi$ involving yet another function $\psi(z)$ to be determined. Inserting $\sig=\bar\sig$, we arrive at
\begin{equation}
  0 = \partial_z(\psi+v\bar\sig) = \partial_zE
\end{equation}
with $E\equiv\psi+v\bar\sig$. Hence, in steady state $E$ has to remain constant throughout even if the strain rate $\dgam$ varies.

To determine the integrating factor, we make the ansatz
\begin{equation}
  \sig\partial_zv = \partial_z\left(\phi - \frac{\kap v'}{2}|\partial_z\dgam|^2\right).
  \label{eq:cr:td}
\end{equation}
Here and in the following, the prime denotes the derivative with respect to $\dgam$. The second term becomes
\begin{equation}
  \partial_z\left(\frac{\kap v'}{2}|\partial_z\dgam|^2\right) = \left(\frac{\kap'v'+\kap v''}{2}|\partial_z\dgam|^2+\kap v'\partial_z^2\dgam\right)\partial_z\dgam
  \label{eq:if}
\end{equation}
exploiting the chain rule. The condition
\begin{equation}
  \kap'v' + \kap v'' = 0
  \label{eq:cond}
\end{equation}
constitutes an ordinary differential equation that determines $v(\dgam)$. Imposing this condition, we find from Eq.~\eqref{eq:cr:td}
\begin{equation}
  \sig = \pd{\phi}{v}-\kap\partial_z^2\dgam, \qquad \sig_0 = \pd{\phi}{v}
  \label{eq:cr:v}
\end{equation}
so that $\phi=\phi(\dgam)$ is a local (point-wise) function of the strain rate. Moreover, from Eq.~\eqref{eq:cr:td} we read off
\begin{equation}
  \psi = \frac{\kap v'}{2}|\partial_z\dgam|^2 - \phi,
\end{equation}
where the first term is reminiscent of a kinetic energy with ``mass'' $m\equiv\kap v'$ and $z$ playing the role of time. Since rearranging Eq.~\eqref{eq:cond} leads to $v''/v'=-\kap'/\kap$ and thus $v'=m/\kap$, where $m$ is the integration constant.

Multiplying Eq.~\eqref{eq:cr:v} by $v'$ and again using $\sig=\bar\sig$ leads to
\begin{equation}
  m\partial_z^2\dgam - \pd{(\phi-v\bar\sig)}{\dgam} = 0,
  \label{eq:el}
\end{equation}
which we recognize as Euler-Lagrange equation for the Lagrangian
\begin{equation}
  L(\dgam,\partial_z\dgam) = \frac{m}{2}|\partial_z\dgam|^2 + \phi - v\bar\sig
  \label{eq:lagrangian}
\end{equation}
with $m'=\kap'v'+\kap v''=0$ [Eq.~\eqref{eq:cond}]. Consequently,
\begin{equation}
  E = \frac{m}{2}|\partial_z\dgam|^2 + U
\end{equation}
corresponds to the conserved total ``energy'' in this language and $U\equiv v\bar\sig-\phi$ is a ``potential energy''. While Eq.~\eqref{eq:cr} has already been interpreted as a dynamical system~\cite{olmsted99}, here we have shown that it can be mapped onto the Euler-Lagrange equation~\eqref{eq:cr:v} through transforming the non-local stress into a functional derivative [Eq.~\eqref{eq:if}], which modifies the interfacial coefficient to a constant $m$.

The derivative of the potential energy $U$ reads
\begin{equation}
  \pd{U}{\dgam} = v'\pd{U}{v} = v'(\bar\sig-\sig_0)
\end{equation}
using Eq.~\eqref{eq:cr:v}. Hence, admissable bulk steady states with $\sig_0=\bar\sig$ correspond to extrema of $U$, and coexisting states further require that $U$ is equal.

\subsection{Linear response}

The Euler-Lagrange equation~\eqref{eq:cr:v} follows from minimizing the action functional
\begin{equation}
  \inf_{\dgam(z)} \Int{z} L(\dgam(z),\partial_z\dgam(z))
  \label{eq:action}
\end{equation}
with $L$ given in Eq.~\eqref{eq:lagrangian}. To obtain some insight into this functional, let us consider a homogeneous system in the limit $v\to\dgam$. The minimization becomes
\begin{equation}
  \inf_{\dgam}[\phi(\dgam) - \dgam\sig],
\end{equation}
from which we immediately recognize Onsager's principle with stress $\sig=\phi'(\dgam)$~\cite{onsager31a}. Strain rate and stress are conjugate variables with respect to $\phi(\dgam)$. In a thermal system, $\phi$ plays the role of a thermodynamic potential (a free energy) and captures the fluctuations of the strain rate. The dissipated heat (per time and volume) is then given through $\dgam\sig$. Outside the linear response regime, the variational principle Eq.~\eqref{eq:action} still holds but the variable conjugate to stress has become $v$.

\subsection{Multiple variables}
\label{sec:mult}

So far we have discussed situations where the strain rate is the sole relevant variable. The generalization to multiple variables is straightforward. We collect these variables into the vector $\vec q=(q_1,\dots)$ and, as before, we consider steady states and assume that translational invariance in broken in one spatial dimension, $\vec q=\vec q(z)$. We now require a constitutive relation for each variable, which we assume to be of the form
\begin{equation}
  \sig_i = \sig_{0,i} - \kap_{ij}\partial_z^2q_j
\end{equation}
with bulk stresses $\sig_{0,i}(\vec q)$ and interfacial coefficients $\kap_{ij}(\vec q)$. Throughout this subsection, we sum over repeated indices. Mechanical equilibrium corresponds to uniform stresses $\sig_i=\bar\sig_i$, which, however, do not need to be equal.

For each variable, we seek a conjugate $v_i=v_i(\vec q)$ so that
\begin{equation}
  \sig_i\partial_zv_i = -\partial_z\psi
  \label{eq:integrability}
\end{equation}
holds with $E=\psi+v_i\bar\sig_i$. Inserting the ansatz
\begin{equation}
  \psi = \frac{1}{2}m_{jk}(\partial_zq_j)(\partial_zq_k) - \phi, \quad
  m_{jk} \equiv \kap_{ij}\pd{v_i}{q_k}
\end{equation}
into Eq.~\eqref{eq:integrability}, we find two integrability conditions: the matrix $m_{jk}$ needs to be symmetric, $m_{kj}=m_{jk}$, and all its entries need to be constant coefficients independent of $\vec q$. We thus have obtained a set of first-order differential equations that determine the conjugate variables $v_i$. Once these have been determined, we construct the potential $U(\vec q)=v_i\bar\sig_i-\phi$ with $\sig_{0,i}=\partial\phi/\partial v_i$. Any steady inhomogeneous state is then determined by the set of conditions
\begin{equation}
  \left.\pd{U}{q_k}\right|_{\vec q_\nu} = 0, \qquad U(\vec q_\nu) = E
\end{equation}
for the coexisting bulk phases $\vec q_\nu$ enumerated by $\nu$.


\section{Illustrations}

\begin{figure*}[t]
  \centering
  \includegraphics{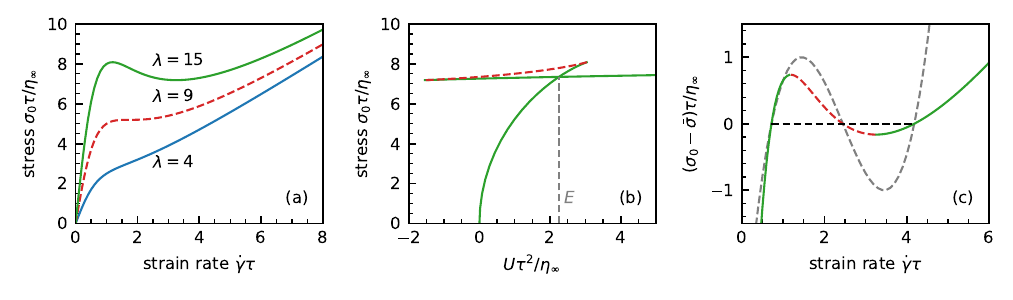}
  \caption{Predicting the steady flow profile. (a)~Bulk shear stress $\sigma_0=\eta\dgam$ from Eq.~\eqref{eq:dhont} as a function of strain rate $\dgam$ for several viscosity contrasts $\lam$. The threshold is $\lam_\text{c}=9$ (dashed line), below the stress is monotonic while for $\lam>9$ an instability is possible. (b)~Parametric plot of stress vs. the effective ``potential energy'' $U=v\sig_0-\phi$ as function of $\dgam$ for $\lam=15$. The red dashed line indicates the unstable branch joining the monotonic low and high stress branches. At the crossing both branches have the same value $E$, which corresponds to the coexistence of two bulk strain rates $\dgam_\pm$. (c)~Zoom into the unstable region for $\lam=15$. The horizontal black dashed line is the uniform stress $\bar\sig$ adopted by the system. The dashed line shows the scaled bulk stress $(\sig_0-\bar\sig)/\kap$, for which coexistence corresponds to an equal area construction [see Eq.~\eqref{eq:maxwell}].}
  \label{fig:illu}
\end{figure*}

\subsection{Shear-thinning fluids}

We return to the strain rate $\dgam$ as the single relevant variable. Following Ref.~\cite{dhont99}, we first consider the simple model expression
\begin{equation}
  \eta(\dgam) = \frac{\eta_0+\eta_\infty(\dgam\tau)^2}{1+(\dgam\tau)^2}
  \label{eq:dhont}
\end{equation}
for the viscosity with bulk stress $\sig_0=\eta\dgam$. This expression interpolates between the zero-shear viscosity $\eta_0$ and the high-shear limit $\eta_\infty$, describing a shear-thinning fluid for $\eta_0>\eta_\infty$. The interpolation is controlled by the time scale $\tau$. The interfacial coefficient is chosen as
\begin{equation}
  \kap(\dgam) = \frac{m}{1+(\dgam\tau)^2}
  \label{eq:kap}
\end{equation}
and monotonically decreases for increasing strain rate.

From $\sig_0'(\dgam)=0$ we find the relation 
\begin{equation}
  [(\dgam\tau)^\text{in}_\pm]^2 = \frac{1}{2}\left(\lam-3\right) \pm \frac{1}{2}\sqrt{\lam^2-10\lam+9}
  \label{eq:spinodal}
\end{equation}
bounding the region within which several strain rates exhibit the same stress. The relevant parameters are the reduced strain rate $\dgam\tau$ and the viscosity contrast $\lam\equiv\eta_0/\eta_\infty$, and above the critical value $\lam_\text{c}=9$ there is a region of mechanical instability within which $\sig'_0<0$ [Fig.~\ref{fig:illu}(a)].

The first step is to determine the function
\begin{equation}
  v(\dgam) = \dgam\left[1 + \frac{1}{3}(\dgam\tau)^2\right],
  \label{eq:v}
\end{equation}
which follows through simple integration of $v'=m/\kap$. We have choosen the integration constants so that $v=\dgam$ coincides with the strain rate in the Newtonian limit $\tau\to0$ (with constant viscosity $\eta=\eta_0$). The second step is to determine the bulk potential
\begin{equation}
  \phi(\dgam) = \IInt{x}{0}{\dgam} v'(x)\eta(x)x
\end{equation}
through integration, which yields the monotonic polynomial
\begin{equation}
  \phi(\dgam) = \frac{1}{2}\eta_\infty\dgam^2\left[\lam+\frac{1}{2}(\dgam\tau)^2\right]
\end{equation}
obeying $\phi(-\dgam)=\phi(\dgam)$. With this result, in Fig.~\ref{fig:illu}(b) we plot $v\sig_0-\phi$ on the $x$-axis and $\sig_0$ on the $y$-axis varying the strain rate. We see that there are two monotonic branches joined by a branch corresponding to the unstable region of strain rates. The crossing of the two stable branches is exactly at coexistence since the two strain rates exhibit the same $U(\dgam_\pm)=E$. In Fig.~\ref{fig:illu}(c) we highlight that the resulting stress $\bar\sig$ does not correspond to an equal-area construction on the bulk stress $\sig_0$, although we note that the shifted and rescaled stress $(\sig_0-\bar\sig)/\kap$ does obeys Eq.~\eqref{eq:maxwell}.


\begin{figure}[b!]
  \includegraphics{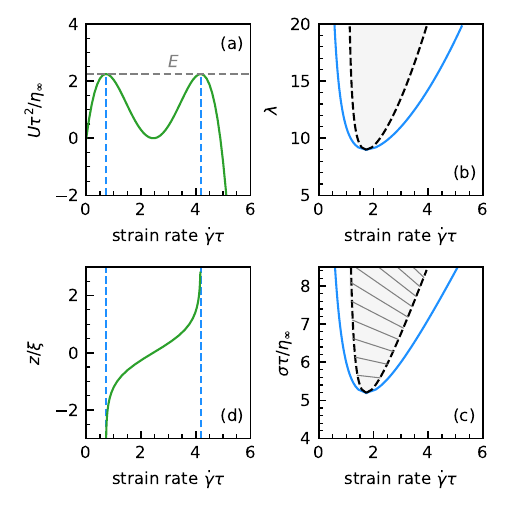}
  \caption{Phase diagram for shear-thinning fluids. (a)~Potential energy $U$ at coexistence. From the maxima we read of the coexisting strain rates (blue dashed lines). (b)~Line of coexisting strain rates [solid blue line, Eq.~\eqref{eq:binodal}] as the viscosity contrast $\lam$ is changed. Dashed lines bound the region within which homogeneous flow is mechanically unstable [Eq.~\eqref{eq:spinodal}]. (c)~In the plane spanned by strain rate and shear stress. We also show tie lines connecting the bulk stresses $\sig_0$ along the dashed line for the same viscosity contrast $\lam$. (d)~Steady flow profile $\dgam(z)$ with the interface centered at $z=0$.}
  \label{fig:phasedia}
\end{figure}

We can now work out the complete phase diagram analytically. This is most easily done through transforming the polynomial $U=v\sig_0-\phi$ into the manifestly symmetric form [Fig.~\ref{fig:phasedia}(a)]
\begin{equation}
  \frac{U\tau^2}{\eta_\infty} = \frac{U(\dgam_0)\tau^2}{\eta_\infty} + \frac{1}{4}[(\lam-\lam_\text{c})-(\delta\dgam\tau)^2](\delta\dgam\tau)^2
  \label{eq:U}
\end{equation}
setting $\dgam=\dgam_0+\delta\dgam$, which requires $\dgam_0=\bar\sig/(3\eta_\infty)$ and
\begin{equation}
  \frac{\bar\sig\tau}{\eta_\infty} = \sqrt{\frac{27}{2}\left(\frac{\lam}{3}-1\right)}.
\end{equation}
The later condition yields the stress adopted by the inhomogeneous flow. The two maxima of $U$ [Eq.~\eqref{eq:U}] determine the coexisting strain rates (also known as binodal)
\begin{equation}
  (\dgam\tau)_\pm = \dgam_0\tau \pm \sqrt\frac{\lam-\lam_\text{c}}{2},
  \label{eq:binodal}
\end{equation}
which we plot in Fig.~\ref{fig:phasedia}(b) as a function of viscosity constrast $\lam$. In Fig.~\ref{fig:phasedia}(c), we also show the phase diagram in the plane spanned by strain and stress.

Finally, the strain rate $\dgam(z)$ is the solution of the Euler-Lagrange equation~\eqref{eq:el} inserting Eq.~\eqref{eq:U},
\begin{equation}
  \frac{m\tau}{\eta_\infty}\partial_z^2(\delta\dgam\tau) + \frac{1}{2}(\lam-\lam_\text{c})\delta\dgam\tau - (\delta\dgam\tau)^3 = 0.
\end{equation}
The solution [Fig.~\ref{fig:phasedia}(d)]
\begin{equation}
  \delta\dgam\tau = \sqrt{\frac{\lam-\lam_\text{c}}{2}}\tanh\frac{z}{\xi}, \qquad
  \xi^2 = \frac{m\tau}{\eta_\infty}\frac{4}{\lam-\lam_\text{c}}
\end{equation}
is well-known and agrees with the binodal [Eq.~\eqref{eq:binodal}] for $z\to\pm\infty$. This solution is translationally invariant and we have placed the center of the interface at $z=0$. The length $\xi$ determines the spatial extent of the interface and diverges as we approach the critical point at $\lam_\text{c}$.

\subsection{Coexistence of solid and sheared liquid}

Another intriguing question concerns the coexistence of a crystalline solid with its melt when shear is applied. Butler and Harrowell have studied a monocomponent Lennard-Jones fluid below the melting temperature at constant pressure and in the presence of explicit walls~\cite{butler02,butler03}. In the quiescent case, the particles arrange into a crystalline structure that spans the system. When applying a (small) steady strain rate $\dgam_\text{a}$ through moving the walls, the system partially melts and the solid coexists with a sheared liquid; comparable to the coexistence of solid and liquid in thermal equilibrium at fixed volume. While Butler and Harrowell make a distinction between the linear response in this case and the emergence of bands due to non-linearities, all observations in their simulations are compatible with the picture of a mechanical phase transition. In particular, varying the applied strain rate the stress and the liquid strain rate remain constant. What changes is the fraction of the liquid.

What are the conditions on the constitutive equation to exhibit coexistence with a solid? First, we make the ansatz
\begin{equation}
  \sig_0 = \sig_\text{eq} + \frac{\al(\dgam-\dgam_0)+\eta_\infty\tau^2(\dgam-\dgam_0)^3}{1+(\dgam\tau)^2}
  \label{eq:coex}
\end{equation}
with two parameters $\al$ and $\dgam_0$, which again reduces to $\sig_0\simeq\eta_\infty\dgam$ in the limit of large strain rate. Here, $\sig_\text{eq}$ is the stress of the equilibrium crystal in the absence of external shear. For the interface, we retain Eq.~\eqref{eq:kap} and, consequently, Eq.~\eqref{eq:v}. The effective potential $U=\bar\sig v-\phi$ then reads
\begin{equation}
  U = (\bar\sig-\sig_\text{eq})v - \frac{\al}{2}(\dgam-\dgam_0)^2 - \frac{\eta_\infty\tau^2}{4}(\dgam-\dgam_0)^4.
\end{equation}
The solid does not flow, which requires $\dgam_-=0$. Moreover, we need to fulfill $U'(\dgam_\pm)=0$ and $U(\dgam_-)=U(\dgam_+)$. All of these conditions are met through choosing $\bar\sig=\sig_\text{eq}$ and $\dgam_0\tau=\sqrt{-\al/\eta_\infty}$ together with $\al<0$. Expanding Eq.~\eqref{eq:coex} to linear order, we find a positive viscosity $\eta_0=-2\al>0$. In agreement with the computational observations, we find that the stress and the strain rate of the melt $\dgam_+=2\dgam_0$ are independent of applied strain $\dgam_\text{a}$, while the lever rule predicts that the fraction $f=\dgam_\text{a}/\dgam_+$ of the system that is liquid grows proportional to the applied strain rate. Once $\dgam_\text{a}$ surpasses $\dgam_+$ there is only sheared liquid ($f=1$), which follows the homogeneous constitutive relation $\sig_0(\dgam)$.

It is instructive to briefly revisit the argument of Ref.~\cite{butler03} against a variational principle. Minimizing the Gibbs free energy implies that the chemical potentials of solid and liquid need to be equal. We know the chemical potential of the solid, which is elevated by the elastic deformation. The observation is that the chemical potential difference between the strained crystal and the liquid at rest remains negative, and that reasonable non-equilibrium generalizations of the liquid chemical potential (e.g., through the Jarzynski relation~\cite{richard19}) only increase the difference. That the Gibbs free energy, and the chemical potential derived from it, is not a good candidate does not preclude that a variational principle exists. As is maybe not surprising, a more fruitful direction is the Onsager principle governing the dissipation in the linear response regime. As shown here, this variational principle can be extended beyond the linear response regime through a generalized conjugated variable $v$ that is determined from an integrating factor.


\section{Conclusions}

To conclude, we have revisited the description of inhomogeneous flows in complex fluids as a phase coexistence from the perspective of a ``mechanical'' phase transition motivated by recent insights into phase equilibria of active particles. The only ingredients are a non-local constitutive relation between strain rate and stress, and the requirement of mechanical equilibrium. Noticeably, no notion of temperature or generalizations of a free energy are required and the framework applies to both (iso)thermal and athermal systems. Cast as an effective dynamical system (with space taking the role of time), the fact that the corresponding energy is a conserved quantity allows us to unambiguously obtain the coexisting strain rates as well as the strain rate profile. As an illustration, we have considered the model viscosity from Ref.~\cite{dhont99} describing a shear-thinning fluid. For a mechanical instability to occur, the fluid needs to be strongly thinning with a viscosity contrast $\lam>9$. For comparison, hard spheres have $\lam\simeq 2$~\cite{squires05} and we would thus not expect to observe shear banding in model hard spheres~\cite{royall24}. We have then argued that there is no fundamental difference between shear bands and the coexistence of a sheared solid with its melt, which fall into the same theoretical framework. Attempts to generalize the free energy to steady states are plagued by inconsistencies, which are avoided from the mechanical perspective followed here.

The basis of our approach is mesoscopic since details of microscopic interactions only appear through the coefficients of the constitutive relation. Clearly, the quality of that relation determines the applicability to real-world systems. To keep illustrations simple, here we have considered the strain rate as the only mechanical variable. However, experiments on, e.g., metallic glasses also show a density variation~\cite{schmidt15}. Other systems require coupling to microstructural quantities such as concentration~\cite{fielding03,vandennoort08,besseling10}. The generalization of the framework is straightforward as demonstrated in Sec.~\ref{sec:mult}. It requires to formulate a constitutive relation for each component, which is left for future work.


\begin{acknowledgments}
  I acknowledge financial support by the Deutsche Forschungsgemeinschaft (DFG) within collaborative research center TRR 146 (Grant No. 404840447).
\end{acknowledgments}


%

\end{document}